\documentclass[aps,pra,reprint,superscriptaddress,showkeys,nofootinbib]{revtex4-2}

\usepackage[dvipsnames]{xcolor}
\usepackage[colorlinks=true,linkcolor=OrangeRed,citecolor=Cerulean,urlcolor=ProcessBlue]{hyperref}
\usepackage{amsmath}
\usepackage{amsfonts}
\usepackage{amssymb}
\usepackage{bbm}
\usepackage{dsfont}
\usepackage{mathrsfs}
\usepackage{array}
\usepackage{dcolumn}
\usepackage{url}
\usepackage{textcomp}
\usepackage{graphicx}
\usepackage{color}
\usepackage{braket}
\usepackage[mathscr]{euscript}
\usepackage{bbm}
\usepackage{enumerate}
\usepackage{appendix}
\usepackage{lipsum}

\usepackage[10pt]{moresize}

\usepackage{float}

\def\alp{\alpha}
\def\bt{\beta}
\def\sg{\sigma}

\def\om{\omega}
\def\Om{\Omega}

\def\lo{\lambda}

\def\Dt{\Delta}

\def\gm{\gamma}

\def\ep{\epsilon}

\def\8{\phi}
\def\7{\Phi}
\def\9{\psi}

\def\fr{\frac}

\def\b{\bra}

\def\kt{\ket}

\def\bk{\braket}
\def\BK{\Braket}
\def\h{\hat}
\def\dg{\dagger}

\def\tens{\otimes}

\def\Am{\mathcal{A}}

\def\Bm{\mathcal{B}}
\def\Cm{\mathcal{C}}
\def\Dm{\mathcal{D}}
\def\Em{\mathcal{E}}

\def\Hm{\mathcal{H}}

\def\Km{\mathcal{K}}

\def\M{\mathcal{M}}
\def\Oe{\mathcal{O}}
\def\Pm{\mathcal{P}}

\def\Sm{\mathcal{S}}
\def\Tm{\mathcal{T}}
\def\Um{\mathcal{U}}
\def\Us{\mathscr{U}}

\def\iD{\mathbbm{1}}
\def\inT{\mathbbm{Z}}

\def\3{\big}
\def\4{\Big}
\def\5{\Bigg}
\definecolor{ogreen}{rgb}{0.2, 0.5, 0.1}
\definecolor{Orange}{rgb}{1, 0.6, 0.4}

\def\ti{\textit}

\def\tt{\text}

\newcommand{\tus}[1]{\underline{\smash{#1}}}

\def\sz{\small}
\def\fsz{\footnotesize}

\newcommand{\tsz}[1]{\text{\small #1}}

\def\nsz{\normalsize}

\def\aref{\autoref}
\newcommand{\appref}[1]{\hyperref[#1]{Appendix~\ref{#1}}}
\newcommand{\apprefNull}[1]{\hyperref[#1]{Appendix}}
\usepackage[english]{babel}
\addto\extrasenglish{%
}

\addto\captionsenglish{}

\def\vo{\vspace{1mm}}
\def\vt{\vspace{2mm}}
\def\vth{\vspace{3mm}}

\def\vTo{\vspace{-1mm}}
\def\vTt{\vspace{-2mm}}
\def\vTh{\vspace{-3mm}}

\def\hs{\kern 0.16667em}
\def\Hs{\kern 0.5em}
\def\hsh{\kern 0.1em}
\def\hshh{\kern 0.07em}
\def\hsn{\kern -0.3em}
\def\hsN{\kern -0.6em}
\def\hseN{\kern -0.55em}
\def\hsnh{\kern -0.15em}
\def\hsnhh{\kern -0.1em}

\newcommand*{\Scale}[2][4]{\scalebox{#1}{$#2$}}%

\newcommand{\SDrop}[2]{\fontdimen16\textfont2=#1
\fontdimen17\textfont2=#1 \tt{$#2$} \fontdimen16\textfont2=2pt
\fontdimen17\textfont2=2pt}
\newcommand{\SRaise}[2]{\fontdimen13\textfont2=#1
\fontdimen14\textfont2=#1 \fontdimen15\textfont2=#1 \tt{$#2$} \fontdimen13\textfont2=2pt
\fontdimen14\textfont2=2pt \fontdimen15\textfont2=2pt}

\usepackage{fancyhdr}
\def\nl{\noindent}

\numberwithin{equation}{section}

\def\tr{\text{Tr}}
	
\def\methodAT{\fsz USER\sz}

\def\methodT{\sz USER\normalsize}
\def\methodTO{\sz USER \normalsize}

\def\aqsT{\sz AQS\nsz}
\def\aqsTO{\sz AQS \nsz}

\def\searT{\sz SEAR\nsz}
\def\searTO{\sz SEAR \nsz}

\usepackage[all]{hypcap}

\usepackage[shortlabels]{enumitem}

\begin{document}

\title{Universalizing Analog Quantum Simulators}
\date{\today}

\author{Andrew Shaw}
\email[Electronic Address: ]{ashaw12@umd.edu}
\affiliation{University of Maryland, College Park, MD 20742, USA}

\begin{abstract}

\ti{Unitary Sampling Expectation-Value Reconstruction} (\methodAT) is used to determine expectation values that are not directly accessible on an analog quantum simulator.
\end{abstract}

\maketitle


\section{Unitary Sampling Expectation-Value Reconstruction}

\ti{Unitary Sampling Expectation-Value Reconstruction} (\methodT), allows the indirect determination of expectation values.

\subsection{Unitary Operators}

For a Hilbert space \tsz{$\Hm$}, the \ti{Haar measure} is comprised of all operators that satisfy unitarity \cite{haar2}:

\begin{align}
\h U \in \Hm\hsh \tens \hsh &\Hm^*,\\[0.4em]
\h U  \hsh \h U^{\dg}=&\h \iD
\end{align}

A unitary operator can be expressed as follows:

\begin{align}
\h U=&\sum_{\alp} \hs \SRaise{5pt}{e^{\Scale[0.8]{\hsh i\8_\alp}}} \kt{\alp}\hsnh \b{\alp}, \\[0.4em]
\8_{\alp}& \in [-\pi,\pi \hsh	]
\end{align}

The \ti{phase separation} is the following:

\begin{align}
\Pm=\4|\8_{\alp}-\8_{\bt}\4|_{\tt{max}}
\end{align}

\subsection{Expectation Values}

A quantum state and an observable can be expressed as follows:

\begin{align}
\kt{\9}=\sum_{\alp} c_{\alp} \kt{\alp} \\[0.4em]
\h O=\sum_{\alp,\bt} \Oe_{\alp,\bt} \kt{\alp}\hsnh \b{\bt}	
\end{align}

The expectation value of the observable after a \ti{similarity transformation} is the following:

\begin{align}
\BK{\9|\h U^{\dg} \hsh \h O \hs \h U|\9} &=\sum_{\alp,\bt} c_{\alp}^{*}c_{\bt} \hsh \Oe_{\alp,\bt} \ e^{i\{\8_{\bt}-\8_{\alp}\}} \\[0.4em]
&=\sum_{\alp,\bt} \Dm_{\alp,\bt} \ \SRaise{5pt}{e^{\Scale[0.8]{\hsh i\hsh \Km_{\alp,\bt}}}}
\end{align}

\begin{figure}
\includegraphics[scale=0.089]{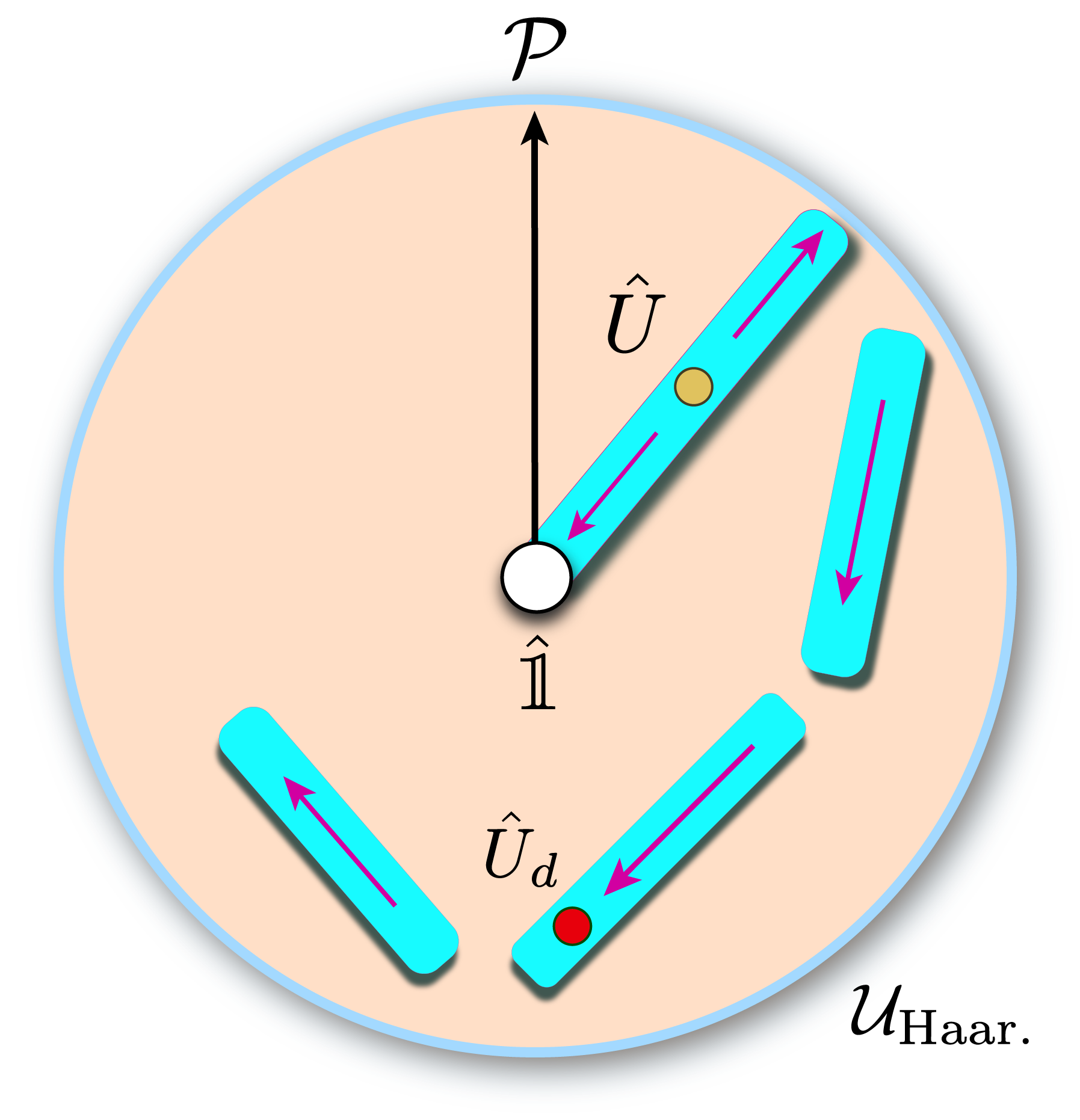}
\caption{ A unitary operator (golden) is exponentiated to generate a multiplicative subset (blue). A discretization unitary (red) is used to reconstruct inaccessible expectation values.}
\label{fig:HaarStructure}
\end{figure}

\subsection{Reconstruction}

A \ti{multiplicative subset} of the Haar measure is obtained by exponentiating a unitary operator (\aref{fig:HaarStructure}):

\begin{align}
\3(\h U\3)^{\eta}=\sum_{\alp}\SRaise{5pt}{e^{\Scale[0.8]{\hsh i\eta \hsh \8_{\alp}}}\kt{\alp}\b{\alp}}
\end{align}

The \ti{multiplicative expectation values} are the following:

\begin{align}
\BK{\h O(\eta)}&=\BK{\9| \3(\h U^{\dg}\3)^{\eta} \ \h O \ \3(\h U\3)^{\eta}|\9} \\[0.6em]
&=\sum_{\alp,\bt} \Dm_{\alp,\bt} \ \SRaise{5pt}{e^{\Scale[0.8]{\hsh i\eta \hsh \Km_{\alp,\bt}}}}	
\end{align}

The multiplicative expectation values can be reconstructed from \ti{discrete multiplicative expectation values}:

\begin{align}
\kern -1.3em \5[\ \BK{\h O\hsh \3(\hsnh -m \hsh \eta_d \hsh\3 )}, \hsh \cdots, \hsh & \BK{\h O\hsh (0)},	 \cdots , \BK{\h O\hsh \3(m \hsh \eta_d\3)} \ \5], \\
& \ \hs m\in \inT \notag
\end{align}

The discrete multiplicative expectation values must be sampled faster than the \ti{aliasing rate} \cite{Nyquist1}:

\begin{align}
\eta_{\tt{alias}}=\fr{\pi}{\ \ \3|\Km_{\alp,\bt}\3|_{\tt{max}}}	
\end{align}

The sampling procedure is accomplished by iteration with a \ti{discretization unitary}:

\begin{align}
\h U_d&=\3(\h U\3)^{\eta_{_d}} \\[0.4em]
\eta_{_d} & \hs \Pm <\pi	
\end{align}

\subsection{Universal Subset}

A \ti{reconstructive Haar subset} can be used to determine the expectation values of the complete Haar measure with \methodT:

\vTh

\begin{equation}
\Um_{\tt{recon.}} \in \  \Um_{\tt{Haar.}}
\end{equation}

\nl
An example reconstructive Haar subset is the following:

\begin{enumerate}
\item The subset contains at least one non-trivial unitary operator from each unique multiplicative subset.
\item \tsz{$\Pm<\pi$} for all such non-trivial unitary operators.
\item All integer powers of the non-trivial unitary operators are present in the subset.
\end{enumerate}

\section{Expanding Analog Quantum Simulators}

\subsection{Simulating Unitary Operators}

\subsubsection{Analog Quantum Simulators}

An \ti{analog quantum simulator} (\aqsT) can perform time evolution for a \ti{Hamiltonian family} \cite{feynmansupremacySim}:

\begin{equation}
\h H_{\gm}(t)=\sum_{\alp} \Em_{\gm,\alp}(t) \hs \Ket{\hsh \alp_{\gm}(t) \hsh}\hsnh \Bra{\hsh \alp_{\gm}(t) \hsh}
\end{equation}

Time evolution simulates the \ti{\aqsTO unitary operators}:

\begin{align}
\raisebox{-0.4em}{\SRaise{6pt}{\h U_{\gm}(t)}}\raisebox{-0.3em}{=} \	\raisebox{-0.4em}{\Scale[1.03]{\Tm}} \ \Scale[0.9]{\5\{} \hs \raisebox{-0.5em}{\Scale[1.1]{e}}^{\hsnh -i \hsh \Scale[1.0]{\Scale[1.2]{\int}_{\hsn 0}^{t} \hs dt^{'} \ \h H_{\gm} \Scale[0.9]{(t^{'})}}}\hs \Scale[0.9]{\5\}}	
\end{align}

\subsubsection{Magnus Expansion}

The \aqsTO unitary operators can be expressed in terms of the \ti{time-independent Hermitian operators}:

\begin{align}
\h U_{\gm}&(t)=\SRaise{5pt}{e^{\Scale[0.9]{\hsh i \h M^{(t)}_{\gm}}}}	 \\[0.6 em]
\h M^{(t)}_{\gm}=&\sum_{\alp}  \hs \M_{\gm,\alp}^{(t)} \hsh \kt{\hsh \alp^{(t)}_{\gm}\hshh }\hsnh \b{\hsh \alp^{(t)}_{\gm}}
\end{align}

The time-independent Hermitian operators can be expressed as a series \cite{MagnusExpansion}:

\begin{align}
\h M^{(t)}_{\gm}=\sum_{n=1}^{\infty}	 \hs \h \Om_{\gm,n}^{(t)}
\end{align}

The first few terms are as follows:

\begin{align}
\h \Om_{\gm,1}^{(t)}&=-\int_0^t \hs dt_1 \hsh \h H_{\gm}(t_1)\\[0.4em]
\h \Om_{\gm,2}^{(t)}&=\fr{i}{2}\int_0^t\hs dt_1 \hsh \int_0^{t_1} \hsh dt_2 \hsh \3[\h H_{\gm}(t_1),\h H_{\gm}(t_2)\3]\\[0.4em]
\cdots\notag
\end{align}

\vTh

\subsubsection{Haar Measure Accessibility}

The \ti{simulated Haar subset} contains the \aqsTO unitary operators:

\vTo

\begin{equation}
\Um_{\tt{sim.}}	\in \  \Um_{\tt{Haar.}}
\end{equation}

\vo

The \ti{expanded Haar subset} contains the unitary operators whose expectation values can be determined with \methodT:

\vTo

\begin{equation}
\Um_{\tt{sim.}}	\in \ \Um_{\tt{exp.}} \in \  \Um_{\tt{Haar.}}
\end{equation}

\subsection{Analog Simulation Expansion}

An \ti{intermediate unitary operator} requires \methodTO to evaluate its expectation values:

\begin{equation}
\begin{split}
\h U_i &\in 	\ \Um_{\tt{exp.}}, \\[-0.1em]
\h U_i	&\notin \ \Um_{\tt{sim.}}
\end{split}
\end{equation}

\vt

\nl
\methodTO is applied in three stages:

\begin{enumerate}[I.]
\item \tus{Exponentiation}: Parametrize the intermediate unitary operator.
\item \tus{Discretization}: Identify an accessible discretization unitary.
\item \tus{Reconstruction}: Reconstruct the intermediate expectation value.
\end{enumerate}

\subsubsection{Exponentiation}

The intermediate unitary operator can be expressed in terms of the \ti{intermediate Hermitian operator}:

\begin{align}
\h U_i=&\SRaise{5pt}{e^{\Scale[0.9]{\hsh i\pi \h A}}} \\[0.4em]	
\h A=\sum_{\alp}& \hs \Am_{\alp} \kt{\alp}\hsnh \b{\alp}, \\[0.1em]
\3|\Am_{\alp}\3| &\leq 1
\end{align}

The \ti{intermediate multiplicative subset} is the following:

\begin{equation}
\3(U_i\3)^{\eta}=\SRaise{5pt}{e^{\Scale[0.9]{\hsh i\pi \eta \hsh \h A}}}
\end{equation}

\subsubsection{Discretization}

The \ti{simulation discretization unitary} is the following:

\begin{align}
&\h U_{s,d} \hsh \in \ \Um_{\tt{sim.}}\\[0.6em]
& \kern -0.7em \h U_{s,d}	=\SRaise{5pt}{e^{\Scale[0.9]{\hsh i \pi \lo \hsh \h A}}}, \\[0em]
& \kern 0.9em \lo < \hs \fr{1}{2}
\end{align}

\subsubsection{Reconstruction}

The \ti{intermediate expectation value} is the following:

\begin{align}
\BK{\h O_i}=\BK{\h U_i^{\dg} \	\hsnh \h O \ \h U_i}
\end{align}

The \ti{minimal intermediate eigenvalue gap} is as follows:

\begin{align}
\Dt \Am_{\tt{min}}	=\4|\hs \Am_{\alp}-\Am_{\bt} \hs \4|_{\tt{min}}
\end{align}

The \ti{discrete sampling unitary operators} are as follows:

\begin{align}
\kern 3em \4[\ \SRaise{6pt}{\h U_{s,d}^{\hshh -n_l}}\hs \Scale[1.3]{,}\ \cdots \ \Scale[1.3]{,} & \ \h \iD \ \Scale[1.3]{,}\	\h U_{s,d} \ \Scale[1.3]{,} \ \cdots \ \Scale[1.3]{,} \ \SRaise{6pt}{\h U_{s,d}^{n_l}} \ \4], \\[0.9em]
 n_l\hsh \gg \ &\fr{2+\Dt \Am_{\tt{min}}}{\lo \hs \Dt \Am_\tt{min}}
\end{align}

The \ti{discrete multiplicative expectation values} are as follows:

\begin{align}
\5[\ \BK{\h O_i\hsh (-n_l \hsh \lo)}, \hsh \cdots, \hsh \BK{\h O_i\hsh (0)},	 \cdots , \BK{\h O_i \hsh (n_l \hsh \lo)} \ \5]
\end{align}

The discrete multiplicative expectation values are used to reconstruct the intermediate expectation value \cite{Nyquist0}:

\begin{align}
\BK{\h O_i}\approx \sum_{k=-n_l}^{n_l} \hs \BK{ \h O_i \hsh (k\lo)} \ \tt{sinc} \5[\fr{1-k\hsh \lo}{\lo}\5]
\end{align}

\section{Simulated Expectation-Value Approximate Reconstruction}

\rhead{III SIMULATED EXPECTATION VALUE APPROXIMATION}

\ti{Simulated Expectation-Value Approximate Reconstruction} (\searT), is a method for estimating the expectation value of an intermediate unitary operator.

\vth

\searTO is applied in three stages:

\begin{enumerate}[I.]
\item \tus{Analog Simulation Expansion}: Apply \methodTO to approximate the intermediate expectation value.
\item \tus{Series Approximation}: Repeat Stage I to generate a series of approximations for the intermediate expectation value.
\item \tus{Error Estimation}: Approximate the average error in the intermediate expectation value.
\end{enumerate}

\subsubsection{Analog Simulation Expansion}

The \ti{\tsz{$\kappa^{\tt{th}}$}-order time-independent Hermitian operators} are the following:

\begin{align}
\h M_{\gm,\kappa}^{(t)}= \sum_{n=1}^\kappa \hs \h \Om_{\gm,n}^{(t)}
\end{align}

A sequence of members of the Hamiltonian family is chosen that satisfies the following condition:

\begin{align}
\sum_{\xi=1}^{n_s} \hsh \h M_{\xi,\kappa}^{(t)}	&= \pi \hsh \lo \hsh \h A, \\[-0.3em]
& \kern -0.92em \lo<\fr{1}{2}
\end{align}

The \ti{approximate simulation discretization unitary} is the following:

\begin{align}
\h U^{'}_{\hsnh s,d}&=\hs \prod_{\xi=1}^{n_s} \hs \SRaise{5pt}{e^{\Scale[0.9]{\hsh i\h M_{\xi}^{(t)}}}} \\[0.4em]
&\approx \SRaise{5pt}{e^{\Scale[0.9]{\hsh i\pi \lo \h A}}}
\end{align}

Performing reconstruction yields the \ti{approximate intermediate expectation value}:

\begin{align}
\BK{\h O_{i,a}}= \BK{\h U_{i,a}^{\dg} \hsnh \ \h O \ \h U_{i,a}}	
\end{align}

\subsubsection{Series Approximation}

\begin{figure}
\includegraphics[scale=0.1]{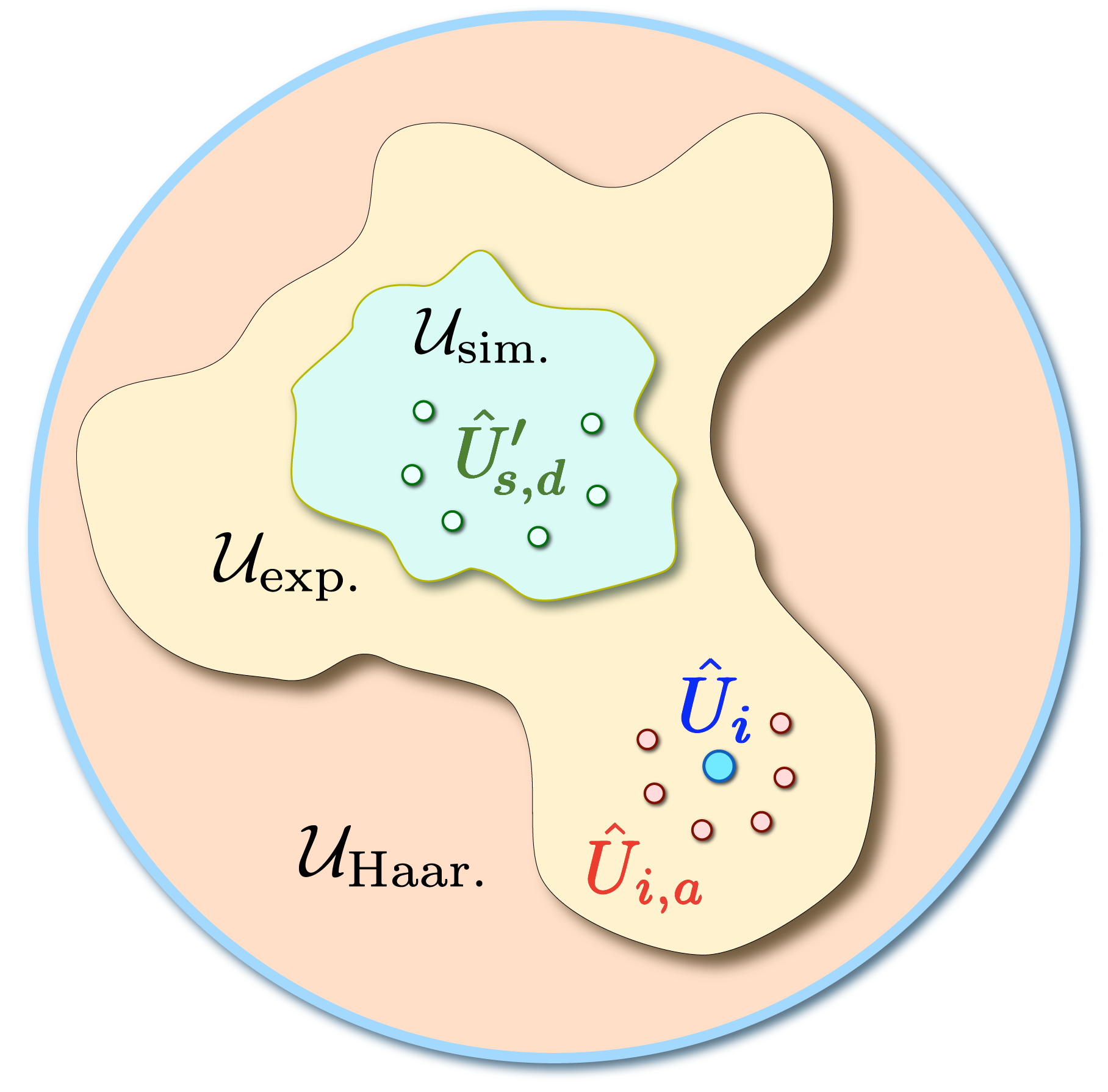}
\caption{Approximate simulation discretization unitaries (green) are used to generate approximate intermediate unitaries (red). These are used to probe the intermediate unitary (blue).}
\label{fig:SearStructure} \vspace{-6mm}
\end{figure}

The expectation values for a series of \ti{approximate intermediate unitaries} are obtained (\aref{fig:SearStructure}):

\begin{align}
\4\{\h U_{i,a}^{(1)} \hsh , \hsh \h U_{i,a}^{(2)} \hs, \hsh \cdots, \hs	U_{i,a}^{(n_a)}\4\}
\end{align}

\subsubsection{Error Estimation}

The \ti{mean approximate intermediate expectation value} is the following:

\begin{equation}
\BK{\h O_{i,a}}_{\tt{mean}}=\fr{1}{n_a}	\hs \sum_{k=1}^{n_a} \ \BK{\h U_{i,a}^{(k)\dg} \hsnh \ \h O \ \h U_{i,a}^{(k)}}
\end{equation}

The error in this quantity can be approximated using \ti{quantum channel technology}.

\subsection{Quantum Channel Technology}

\subsubsection{Density Matrix Formalism}

A \ti{density matrix} describes an ensemble of quantum states \tsz{$\kt{\9_k}$}, each with observational probability \tsz{$p_k$} \cite{densmat1}:

\begin{equation}
\rho=\sum_{k} p_k \kt{\9_k}\hsnh \b{\9_k}	
\end{equation}

Density matrices satisfy the following condition:

\begin{equation}
\tr\3[\hsh \rho \hsh \3]=1	
\end{equation}

Expectation values of density matrices are as follows:

\begin{equation}
\bk{\h O}=\tr\3[\hsh \rho \hsh \h O \hsh \3]	
\end{equation}

\subsubsection{Quantum Channel Formalism}

\ti{Quantum channels} are a type of \ti{superoperator} \cite{superOp}:

\begin{equation}
\h \Cm=\sum_{\mu} \h K_{\mu} \tens \h K_{\mu}^{\dg}	
\end{equation}

They satisfy the following condition:

\begin{equation}
\sum_{\mu} \h K_{\mu} \h K_{\mu}^{\dg}=\h \iD
\end{equation}

Quantum channels act on density matrices as follows:

\begin{align}
\rho'&=\h \Cm \hsh [\hsh \rho \hsh] \\
&=\sum_{\mu} \h K_{\mu}\hs  \rho \hs \h K_{\mu}^{\dg}	
\end{align}

\vTh

\subsection{SEAR Error Channel}

The intermediate expectation value is the following:

\begin{align}
\BK{\h O_i}&=\tr\4[\hs \rho_i \ \h O \hs \4]\\[0.7em]
\rho_i&=\h U_i \ \rho \ \h U_i^{\dg}
\end{align}

The mean approximate intermediate expectation value is as follows:

\begin{align}
\BK{\h O_{i,a}}_{\tt{mean}}	=\fr{1}{n_a} \ \sum_{k=1}^{n_a} \tr\5[\hs \h U^{(k)}_{i,a} \ \rho \ \h U^{(k)\dg}_{i,a} \ \h O \hs \5]
\end{align}

This quantity can be written using the \ti{\searTO error channel}:

\begin{align}
& \kern -4em \BK{\h O_{i,a}}_{\tt{mean}}=\tr\4[	\ \h \Sm \hsh\3[\hsh \rho_i \hsh\3] \ \h O \ \4] \\[0.6em]
\h \Sm &=\sum_{\mu=1}^{n_a} \h S_{\mu} \tens \h S_{\mu}^{\dg}	\\[0.4em]
\h S_{\mu}&=\fr{1}{n_a} \hs \h U_{i,a}^{(\mu)} \hs \h U_i^{\dg}
\end{align}

\subsection{SEAR Error}

\subsubsection{Quantum Channel Averaging}

Applying a similarity transformation to a quantum channel yields the following:

\begin{align}
\h \Cm_u&= \h \Us^{\dg} \ \h \Cm \ \hsh \h \Us \\[0.4em]
&=\sum_{\mu} \h U^{\dg} \hsh \h K_{\mu} \hsh \h U \hs \tens \hs \h U^{\dg} \hsh \h K^{\dg}_{\mu} \hsh \h U
\end{align}

Integrating over the Haar measure yields a \ti{depolarizing channel} \cite{gwnm1}:

\begin{align}
\h \Dm_{\Scale[0.92]{\ep}}=\int	d\hsh \Um_{\tt{Haar}} \ \hs \h \Us^{\dg} \ \h \Cm \ \hsh \h \Us
\end{align}

Depolarizing channels mix quantum states with the identity at \ti{noise strength} \tsz{$\ep$}:

\begin{align}
\h \Dm_{\Scale[0.92]{\ep}}\hsh \3[\hsh \rho \hsh\3]=\3(\hsh 1-\ep \hsh \3) \hsh \rho \ +\ \ep \hsh \fr{\iD}{\tt{dim}(\Hm)} 
\end{align}

The \ti{\searTO depolarizing channel} is  the following:

\begin{align}
\h \Dm_{\Scale[0.92]{\ep_s}}=\int d \hsh \Um_{\tt{Haar}} \ \hs \h \Us^{\dg} \ \h \Sm	\ \hsh \h \Us
\end{align}

\subsubsection{Expectation Value Error}

The \ti{intermediate expectation value error} is as follows:

\begin{align}
\Dt O^{i}=\5| \ \tr \5[\hs \4(\hsh \rho_i -\Sm \hsh\3[\hsh \rho_i \hsh\3] \hsh \4) \ \h O \hs \5]\ \5|
\end{align}

The \ti{average intermediate expectation value error} is the following:

\begin{align}
\kern -1em \Dt O^{i}_{\tt{mean}}	&=\hs \5|\ \tr\5[\hs \4(\hsh \rho_i-\int d \hsh \Um_{\tt{Haar}} \ \Sm_u \hsh \3[\hsh \rho_i \hsh\3]\hsh \4) \ \h O \hs \5] \ \5| \\[0.6em]
&=\hs \5|\ \tr\5[\hs \4(\hsh \rho_i-\ \h \Dm_{\Scale[0.92]{\ep_s}}\hsh\3[\hsh\rho_i\hsh\3] \hsh \4)\ \h O \hs \5] \ \5|
\end{align}

It can be expressed using the \ti{\searTO noise strength}:

\begin{align}
\Dt O^{i}_{\tt{mean}}=	\ep_s \ \5| \ \BK{\h O_i}-\fr{\tr\3[\h O\3]}{\tt{dim}(\Hm)}\ \5|
\end{align}

\subsection{Estimating the SEAR Error}

\subsubsection{Complementary \searTO Error Channel}

The \ti{complementary \searTO error channels} are the following:

\begin{align}
\h \Sm^{(k)}&=\sum_{\mu=1}^{n_a} \hs \h S^{(k)}_{\mu} \hs \tens \hs  \h S_{\mu}^{(k)\dg} \\[0.6em]
\h S_{\mu}^{(k)}&= \fr{1}{n_a} \hs \h U_{i,a}^{(\mu)} \ \h U_{i,a}^{(k)\dg}
\end{align}

The \ti{complementary \searTO depolarizing channels} are as follows:

\begin{align}
\h \Dm_{\Scale[0.92]{\ep_s}}^{(k)}=\int d \hsh \Um_{\tt{Haar}}	\ \hs & \h \Us^{\dg} \ \h \Sm^{(k)} \ \h \Us \\[0.6em]
\ep_s^{(k)}& \approx \ep_s
\end{align}

\subsubsection{Discrete Quantum Channel Averaging}

The \ti{approximate complementary \searTO depolarizing channels} are the following:

\begin{align}
\h \Bm_{\Scale[0.92]{\ep_s}}^{(k)} =\fr{1}{n_t} \ \sum_{m=1}^{n_t} & \ \hs \h \Us^{\dg}_m \ \hsh \h \Sm^{(k)} \ \hsh \h \Us_m \\[0.6em]
\h U_m & \in \ \Um_{\tt{sim.}}
\end{align}

Shown explicitly:

\vTt

\begin{align}
\h \Bm_{\Scale[0.92]{\ep_s}}^{(k)}=\sum_{\mu=1}^{n_a}\sum_{m=1}^{n_t} \hs \h B^{(k)}_{\mu,m}\hs \tens \hs \h B^{(k)\dg}_{\mu,m} 
\end{align}

\vTo

\begin{align}
\h B^{(k)}_{\mu,m}&=\fr{1}{n_a n_t} \ \h U_m^{\dg} \ \h U_{i,a}^{(\mu)} \ \h U_{i,a}^{(k)\dg} \ \h U_m \\[0.6em]
&=	\fr{1}{n_a n_t} \ \h U^{(k)}_{\mu,m}
\end{align}

\subsubsection{Evaluating the Approximate Complementary \searTO Depolarizing Channels }

The \ti{approximate complementary \searTO depolarizing expectation value} is the following:

\begin{align}
\kern -0.4em\BK{\h O^{(k)}}_{\tt{comp.}}&= \tr\5[\ \h \Bm_{\Scale[0.92]{\ep_s}}^{(k)}\3[\hsh\rho\hsh\3] \ \h O\ \5]
\end{align}


This quantity can be expressed using the \ti{partial \searTO depolarizing expectation values}:

\begin{align}
\kern -1.2em\BK{\h O^{(k)}}_{\tt{comp.}}&=\fr{1}{n_a n_t} \ \sum_{\mu,m=1}^{n_a,n_t} \ \BK{\h O^{(k)}_{\mu,m}} \\[0.6em]
&=\fr{1}{n_a n_t} \ \sum_{\mu,m=1}^{n_a,n_t} \tr \5[\ \h U^{(k)}_{\mu,m} \ \rho \ \h U^{(k)\dg}_{\mu,m} \  \h O\ \5]
\end{align}

%
%

\subsubsection{Estimating the Approximate Complementary \searTO Depolarizing Expectation Value}

The \ti{discrete approximate intermediate unitaries} are the following:

\vTh

\begin{align}
\h U^{(k)}_{i,a,d}=\4[ \hs \hsh \h U_{\hsnh s,d}^{'(k)} \hs \4]^{\tau^{(k)}} \\[0.4em]
\tau^{(k)}=\tt{round}\5[ \hs \fr{1}{\lo^{(k)}}\hs  \5]
\end{align}

The \ti{approximate partial \searTO depolarizing expectation values} are the following:

\begin{align}
\BK{\h O^{(k)}_{a,\mu,m}}=&\tr\5[ \ \h U^{(k)}_{a,\mu,m} \ \rho \ \h U^{(k)\dg}_{a,\mu,m} \ \h O \ \5] \\[0.7em]
\h U_{a,\mu,m}^{(k)}&=\h U_m^{\dg} \  \h U_{i,a,d}^{(\mu)} \ \h U_{i,a,d}^{(k)\dg} \ \h U_m
\end{align}

\vt

The \ti{discrete approximate complementary \searTO depolarizing expectation value} is the following:

\begin{align}
\BK{\h O_d^{(k)}}_{\tt{comp.}}	&=\fr{1}{n_a n_t} \ \sum_{\mu,m=1}^{n_a, n_t} \ \BK{\h O^{(k)}_{a,\mu,m}}
\end{align}

\subsubsection{Estimating the \searTO Noise Strength}

The \ti{discrete complementary \searTO noise strength} is as follows:

\begin{align}
\kern -0.7em \ep_{s,d}^{(k)}	=\5[\hs \BK{\h O_d^{(k)}}_{\tt{comp.}}-\tr\4[\hs \rho \ \h O \hs\4]& \hs \5 ]\hs  \5[\ \fr{\tr\3[\h O\3]}{\tt{dim}(\Hm)}-\tr\4[\hs \rho \ \h O \hs \4] \ \5 ]^{-1} \raisetag{-0.6em} \\[0.4em]
\ep_{s,d}^{(k)}&\approx \ep_s^{(k)}
\end{align}

The \ti{mean discrete complementary \searTO noise strength} is the following:

\begin{align}
\overline{ \ep}_{s,d}	=\fr{1}{n_a} \ \sum_{k=1}^{n_a} \hs \ep_{s,d}^{(k)}
\end{align}

\subsubsection{Estimating the Expectation Value Error}

The average intermediate expectation value error can be approximated as follows:

\vTo

\begin{align}
\Dt O^i_{\tt{mean}}\approx \ \overline{\ep}_{s,d}	 \ \5| \ \BK{\h O_i} - \fr{\tr \3[\h O\3]}{\tt{dim}(\Hm)} \ \5|
\end{align}

The observable is the following:

\vTt

\begin{align}
\h O=\sum_{\sg} \hs \SDrop{3pt}{\om_{\sg}} \kt{\SDrop{3pt}{o_{\sg}}}\hsnh \b{\SDrop{3pt}{o_{\sg}}}	
\end{align}

The \ti{observable eigenvalue spread} is the following:

\vTt

\begin{align}
\Dt\om=\4|\ \om_{\tt{max}}-\om_{\tt{min}}\ \4|	
\end{align}

The \ti{approximate average intermediate expectation value error} is as follows:

\vTo

\begin{align}
\Dt O^i_{a,\tt{mean}}= \ \overline{\ep}_{s,d} \ \Dt \om
\end{align}

\subsection{\searTO Result}

\searTO gives the following estimate for the intermediate expectation value:

\begin{align}
\BK{\h O_i}\ \sim \ \BK{\h O_{i,a}}_\tt{mean}\hs  \pm \hs \Dt O^i_{a,\tt{mean}}
\end{align}

\rhead{IV NUMERICAL IMPLEMENTATION}

\section{Numerical Implementation}

The Hamiltonian family is the following:

\vTo

\begin{align}
\h H_{\gm}(t)=\hs -\fr{1}{2m} \hsh \5[\hs \fr{e^{-i\h p a}-e^{i\h p a}}{2a} \hs \5 ]^2	\hs + \hs a \hsh \h x \hs \sin{\3(\om_{\gm} \hsh t\3)}
\end{align}

\searTO is used to perform quantum simulation of the \ti{target Hamiltonian}:

\vTo

\begin{align}
\h H_t=\hs -\fr{1}{2m} \hsh \5[ \hs \fr{e^{-i \h p a}-e^{i\h p a}}{2a} \hs \5]^2 \hs +\hs b \hsh \h x \hs+ \hs \Km \3(\hsh \h p \hsh \3)
\end{align}

\subsubsection{Unitary Sampling}

The \aqsTO is used to compute the discrete multiplicative expectation values (\aref{fig:EncountTimes}).

\subsubsection{Reconstruction}

\begin{figure}[H]
\includegraphics[scale=0.15]{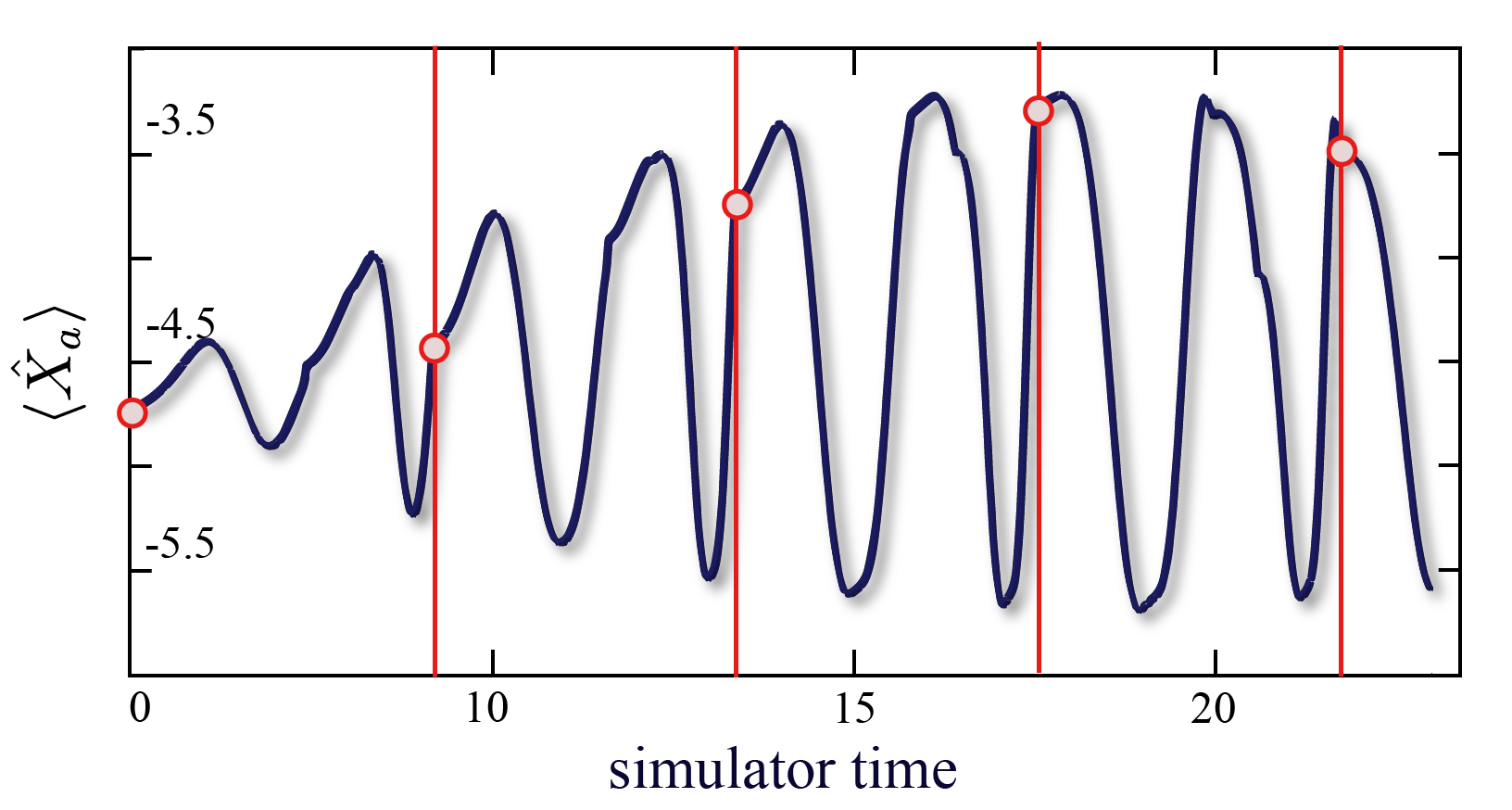}\vspace{-3mm}
\caption{The AQS is used to perform quantum simulation (black). The observable is computed at the \ti{encountering times} (red).}
\label{fig:EncountTimes}

\vth

\includegraphics[scale=0.20]{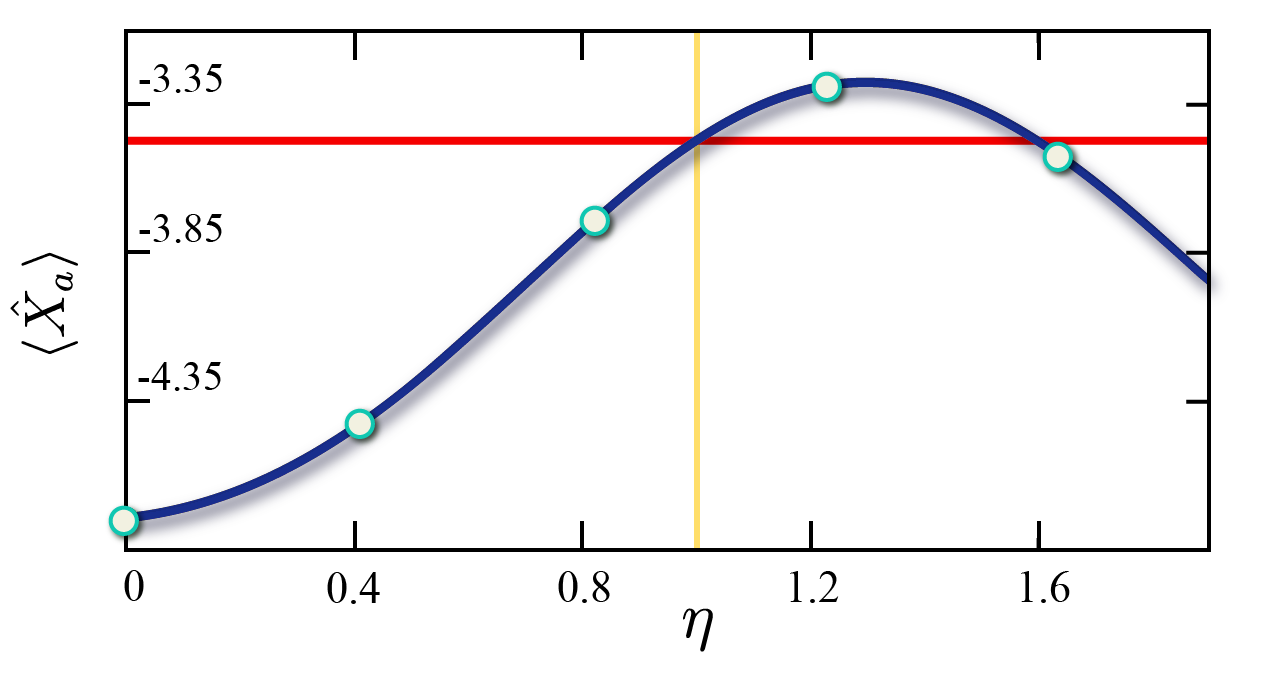}\vspace{-5mm}
\caption{The discrete multiplicative expectation values (green) are reconstructed (blue). The approximate intermediate expectation value (red) is recovered (yellow).}
\label{fig:ReconstructData}\vspace{-4mm}

\end{figure}

The approximate intermediate expectation values are reconstructed (\aref{fig:ReconstructData}).

\subsubsection{Noise Strength Estimation}

The \aqsTO is used to estimate the \searTO noise strength (\aref{fig:EpsilonData}).

\vTh

\subsubsection{SEAR Result}

\begin{figure}[H]
\includegraphics[scale=0.148]{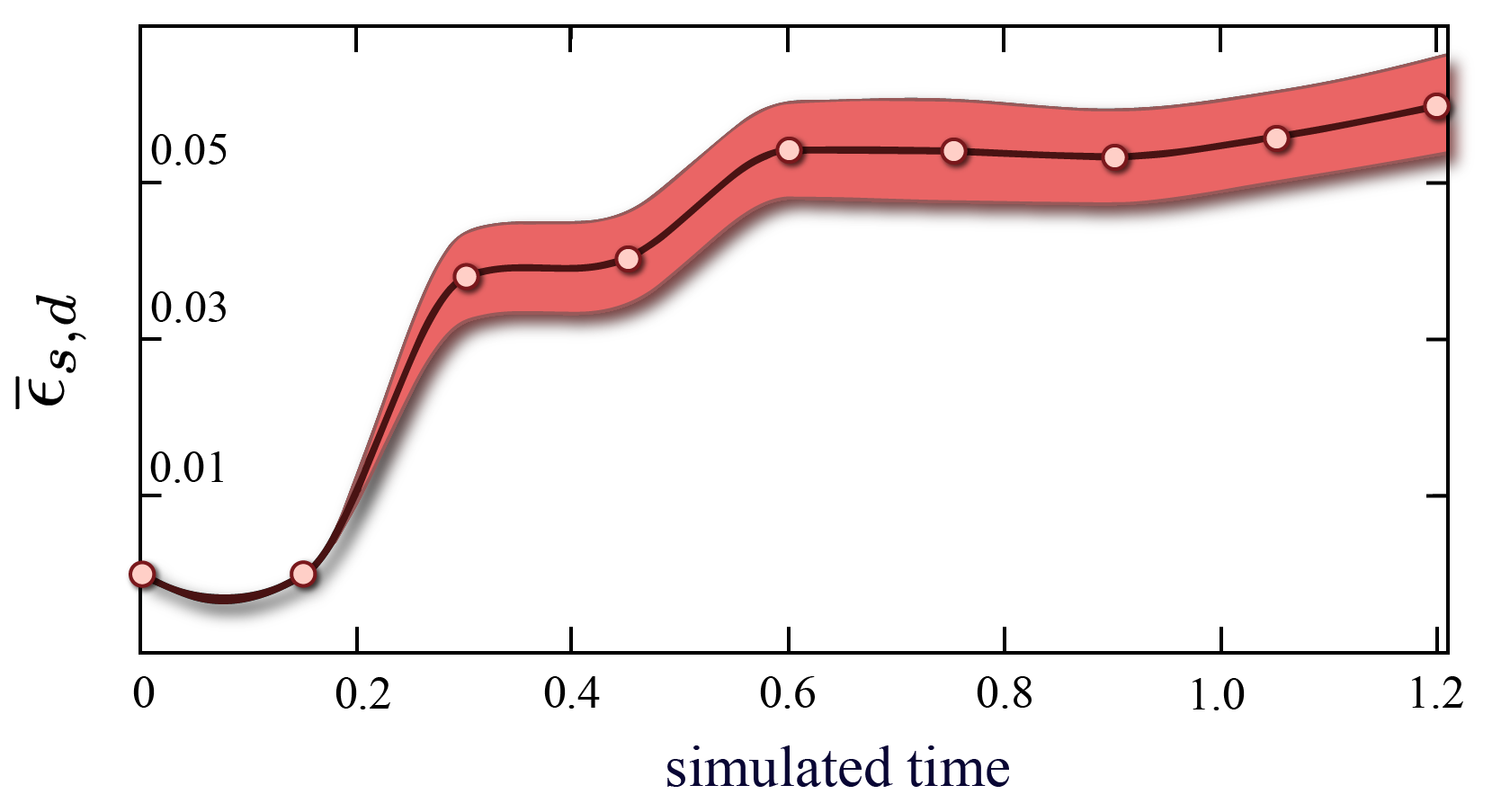}\vspace{-3mm}
\caption{The SEAR noise strength estimate is computed (red).}
\label{fig:EpsilonData}

\vspace{7mm}

\includegraphics[scale=0.148]{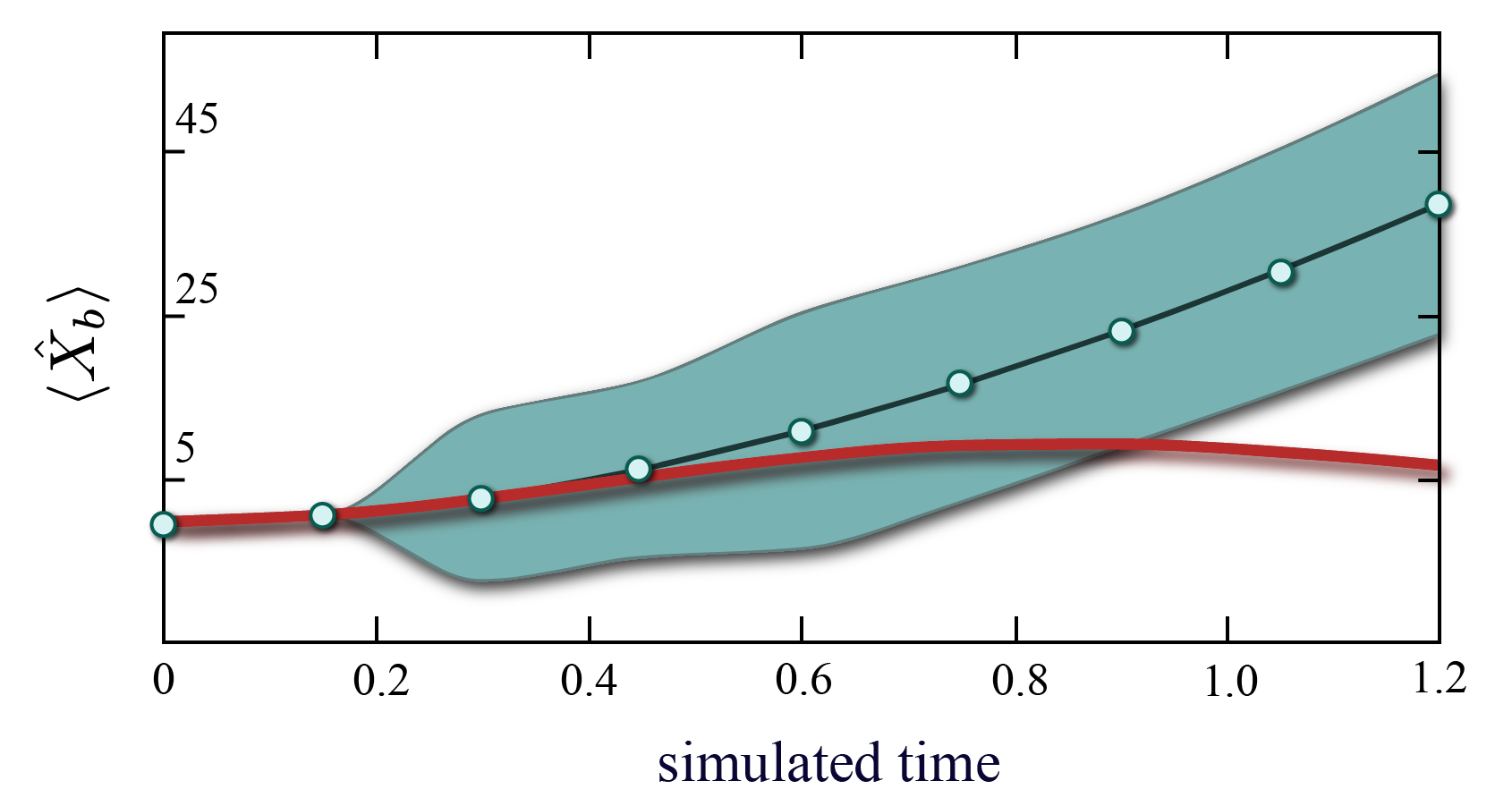}\vspace{-3mm}
\caption{Ideal quantum simulation (red) is contrasted with the SEAR result (aquamarine).}
\label{fig:SearData} \vspace{-4mm}
\end{figure}

The \searTO observable estimate is computed (\aref{fig:SearData}).

\vTh

\section{Appendix}

If the approximate intermediate unitaries meet certain criteria, they can be used to recover the intermediate expectation value \cite{CLAWEArxiv}.

\newpage

\rhead{}

\section{Acknowledgements}\label{sec:ack}

\vTh

\begin{center}

\noindent\rule{3cm}{0.4pt}

\end{center}

\begin{center}
\ti{He was oppressed and treated harshly,  \newline
yet he never said a word.}

\vo
\ti{He was led like a lamb to the slaughter.}

\vo
\ti{And as a sheep is silent before the shearers,\newline
he did not open his mouth.}

\vo

-\ti{Isaiah 53:7}
\end{center}

\begin{center}
$ -AMDG - $
\end{center}

\vspace{5mm}

\renewcommand*{\bibfont}{\scriptsize}

\bibliography{ANAref}

\end{document}